\documentstyle[amssymb,aps]{revtex}

\begin{document}

\noindent \noindent \noindent {\bf  OPTIC-MECHANICAL ANALOGY AND
GRAVITATIONAL EFFECTS IN THE EXTENDED SPACE MODEL}

\bigskip

\bigskip \noindent \noindent {$^1$ V.A.Andreev(*),\\ $^2$D.Yu.Tsipenyuk(**)}

\noindent {$ ^1$Lebedev Physics Institute of Russian Academy of Sciences,
Moscow, Russia}

\noindent {$^2$Prokhorov General Physics Institute of Russian Academy of
Sciences, Moscow, Russia}

\medskip

\noindent

\noindent

\noindent

\noindent \noindent July 2004

\bigskip \noindent (*) andrvlad@yandex.ru

\bigskip \noindent (**) tsip@kapella.gpi.ru

\medskip

\noindent

\begin{center}
\bigskip
\end{center}

{\bf Abstract}

\bigskip

New approach to the description of a gravitation using analogy between the
optical and mechanical phenomenon is advanced. For this purpose in extended
(1+4)-dimensional space $G(1,4)$ the gravitational effects are
considered: the speed of escape, red shift, radar echo, deviation light and
perihelion precession of Mercury.
It is shown, that methods of the Extended Space Model (ESM) give the same
outcomes, as General Theory of Relativity (GR).
The various ways of introduction of refraction index are discussed,
which appropriate to the gravitational field.

\section{\protect\bigskip Introduction}

It is known, that between the mechanical and optical phenomena there is a
certain likeness, which historically was exhibited that a set of the optical
phenomena managed uniformly well to be described both within the framework of
wave, and within the framework of the corpuscular theories. In particular,
motion of a beam of light in an inhomogeneous medium in many respects similar
to motion of a material particle in a potential field [1]. In the given
activity, we shall take advantage of this connection to describe the
gravitational phenomena. Will be shown, that it is possible to receive such
effects of General Theory of Relativity (GR), as red shift, radar echo,
deviation of light and a perihelion precession of a Mercury, using ideas and
methods of geometric optics.

In the papers [2-7] the Extended Space Model (ESM) and the electrodynamics in this space was constructed.
ESM are by generalization of a Special Theory of Relativity (GR) on
(1 + 4) - dimensional space which having the metric (+ - - - -).
We designate it as $ G (1,4) $.
Space of the Minkowski $ M (1,3) $ is a subspace of the ESM.
The role of the fifth coordinates in space $ G (1,4) $ plays an interval in
space of the Minkowski $ M (1,3) $. We shall designate its by letter $ S $.

One of the characteristics of this theory is that in it the rest mass of
particles - is a variable and a photon, falling in the medium with the
refraction index $ n> 1 $, acquires nonzero mass.

The probability that a photon has nonzero mass is widely discussed both
theorists, and experimenters. The review of the last outcomes is contained
in [8]. Our approach differs by that in ESM mass of a particle is not
constant, and is determined by external effects, which it experiences,
theme processes, in which it participates.

According to philosophy of extended space external effects on any object are
described as change of refraction index $ n $ in a point, where there is a
given object. Formally such processes within the framework of our model are
described by rotations in extended space $ G (1,4) $ [2,6].
One of the main physical problems in ESM is how to compare particular
interactions with appropriated to them distributions of an index of
refraction. In each separate case, this problem is decided in its own way.

Gravitational field is one of examples of physical objects, to which compares
some refraction index. From the very beginning origins, GTR there was a
problem of its experimental checking. The majority of observable
gravitational effects connected with the optical phenomena and with a
behavior of photons in a gravitational field. One of such effects is, in
particular - deviation of light in a gravitational field. This deviation can
be interpreted as motion of a light ray in environment with an inhomogeneous
refraction index. Thus, we can to compare to a gravitational field some index
of refraction.

In this paper, we will consider well-known gravitational effects, used for
the GR confirmation. In addition, we shall show, that all of them can be
described and in frameworks ESM.
Let's mark also, that now attempts in a new way to interpret the gravitational
effects are considered by other authors [9-13].

\section{ESM Formalism}

In the Minkowski space $ M (1,3) $ to each particle the 4-vector
energy-momentum is compared [14]
\begin{equation}
{\bf p}\;=\;(\frac Ec,p_x,p_y,p_z).
\label{bb1a}
\end{equation}
In the extended space $ G (1,4) $ we building it up to 5-vector

\begin{equation}
{\bf p}\;=\;(\frac Ec,p_x,p_y,p_z,mc).
\label{bb2a}
\end{equation}

For free particles the components of vector (2) satisfy to an equation
\begin{equation}
E^2 \;=c^2p^2_x+c^2p^2_y+ c ^ 2p^2_z \; + \; m ^ 2c ^ 4 \;,
\label{bb3a}
\end{equation}
i.e. this vector is isotropic.

Parameter $ n $ links speed of light in vacuum $ c $ with speed of light in
the medium $ v = c/n. $ By the help of it is possible to define fifth
coordinate in space $ G (1,4) $.
Thus the empty Minkowski space $ M (1,3) $ corresponds to $ n = 1 $.
In this medium light is gone with speed $ c $. Hit of light in medium with
$ n\ne1 $ is interpreted as an exit of a photon from the Minkowski space and
transition of light in other subspace of space $ G (1,4) $.
Such transition can be described with the help of rotations in space $G(1,4)$.
All types of such rotations are investigated in [2,6].

In blank space in a fixed reference system there are two types of various
object, with zero and nonzero masses.
In space $ G (1,4) $ to them there are corresponds 5-vectors
\begin{equation}
\left(\;\frac{\hbar\omega}c\;,\;\frac{\hbar\omega}c\;,\;0\;\right)\;.
\label{bb4a}
\end{equation}
\begin{equation}
\left(\;mc\;,\;0\;,\;mc\;\right)\;.
\label{bb5a}
\end{equation}
For simplicity we have recorded vectors (4), (5) in (1 + 2)-dimensional space.
The vector (4) describes a photon with zero mass, with energy $\hbar\omega $
and with speed $ c $. The vector (5) describes a fixed particle with weight
$ m $.

At hyperbolic rotations on an angle $ \theta $ in the plane (TS) the photon
vector (4) will be transformed as follows [2,6]

\begin{equation}
\left(\frac{\hbar\omega}c,\frac{\hbar\omega}c,0\right)\; \to\;
\left(\frac{\hbar\omega}c\cosh\theta,\frac{\hbar\omega}c,
 \frac{\hbar\omega}c\sinh\theta\right)=
\left(\frac{\hbar\omega}cn,\frac{\hbar\omega}c,
 \frac{\hbar\omega}c\sqrt{n^2-1}\right).
\label{bb6a}
\end{equation}

Because of such transformations there is a particle with mass

\begin{equation}
m\;=\;\frac{\hbar\omega}{c^2}\sinh\theta\;=\;
\frac{\hbar\omega}{c^2}\sqrt{n^2-1}.
\label{bb7a}
\end{equation}
At these rotations the massive vector (5) will be transformed as follows
\begin{equation}
\left(\;mc\;,\;0\;,\;mc\;\right)\;\;\to\;\;
 \left(\;mce^{\theta}\;,\;0\;,\;mce^{\theta}\;\right),\;\;\;
e^{\theta_{\pm}}\;=\;n \pm \sqrt{n^2-1}  .
\label{bb8a}
\end{equation}

At such rotation  the massive particle changes mass
\begin{equation}
m\;\;\to\;\;m   e^{\theta}\;,\;\;\;0\;\le\;\theta\;<\;\infty\;
\label{bb9a}
\end{equation}
and energy, but is saved a momentum.

At a rotation  on an angle $ \psi $ in a plane (XS) photon vector (4) will be transformed under the law

\begin{equation}
\left(\frac{\hbar\omega}c,\frac{\hbar\omega}c,0\right)
\to\left(\frac{\hbar\omega}c,\frac{\hbar\omega}c\cos\psi,
\frac{\hbar\omega}c\sin\psi\right)=
\left(\frac{\hbar\omega}c,\frac{\hbar\omega}{cn},
\frac{\hbar\omega}{cn}\sqrt{n^2-1}\right).
\label{bb10a}
\end{equation}
Thus the photon acquires mass
\begin{equation}
 m\;=\;\frac{h\omega}{c^2}\sin\psi\;=\;\frac{\hbar\omega}{c^2n}
\label{bb11a}
\end{equation}
and velocity
\begin{equation}
v\;=\;c\cos\psi\;=\;\frac cn.
\label{bb12a}
\end{equation}

The vector (5) massive particles will be transformed under the law
\begin{equation}
\left(mc,0,mc \right)\;\to\;
\left(mc,-mc\sin\psi,mc\cos\psi \right)=
 \left(mc,-\frac{mc}n\sqrt{n^2-1},\frac{mc}n \right).
\label{bb13a}
\end{equation}
Energy of a particle at such transformation is saved, but varies mass
\begin{equation}
M \; \;\to \; \; m\cos\psi \; = \;\frac {m} n
\label{bb14a}
\end{equation}
and momentum

\begin{equation}
0\;\to\;\; -mc\sin\psi\;=\;-\frac{mc}n\sqrt{n^2-1}\; .
\label{bb15a}
\end{equation}

The important property of transformations (6) (10) is that mass of a photon,
which it generates, can have both positive and negative sign. It follows
immediately from properties of a symmetry of space $ G (1,4) $.
As to particles, which initially had positive mass, after transformations (8),
(13) it remains positive.

\vspace{2mm}
\section{Refraction index of a gravitational field}
\vspace{1.em}

Let's study now a problem of refraction index of a gravitational field.
Let there is a dot mass, which gravitational field is described by the
Schwarzchild solution. We assume, that the gravitational radius $ r _ g $ is
small and we shall consider all effects on distances $ r> r _ g $.

In the literature there are two expressions for an index of refraction $ n $,
appropriate to a Schwarzchild field. One of them, we shall name it $ n _ 1 $,
is used in papers of the Okun [15,16] and looks like

\begin{equation}
n_1(r)\;=\;(g_{00})^{-1}\;=\;(1-\frac{r_g}r)^{-1}\;\approx\;1+\frac{r_g}r=
1+\frac{2\gamma M}{rc^2}.
\label{bb16a}
\end{equation}
It is received in the supposition, that in a constant gravitational field the
frequency of a photon $ \omega $ remains to a constant, but the wavelength
$ \lambda $ and speed $ v $ varie.
Other index of refraction $ n _ 2 $ is possible to receive from the formula
of an interval in a weak gravitational field [14]
\begin{equation}
Ds ^ 2 \; = \; (c ^ 2 + 2\varphi) dt ^ 2 - {\bf dr} ^ 2,
\label{bb17a}
\end{equation}
Where $ \varphi $ - is a potential of a gravitational field.
\noindent
Supposing $ \bf {d} r = {\bf v} $ $ dt $ and $ ds ^ 2 = 0 $, we shall receive
speed of a photon in the gravitational field
\begin{equation}
v\;=\;c \left(1+\frac{2\varphi}{c^2}\right)^{1/2}\;
\approx\;c\left(1+\frac{\varphi}{c^2}\right).
\label{bb18a}
\end{equation}
Here it is necessary to take into account that a potential of a gravitational field $ \varphi $ - is negative. For a point source of mass $ M $ we have
\begin{equation}
\varphi (r) \; = \; -\frac {\gamma M} r.
\label{bb19a}
\end{equation}
Substituting expression (19) in the formula (18), we receive

\begin{equation}
v\;\approx\;c\left(1-\frac{\gamma M}{rc^2}\right).
\label{bb20a}
\end{equation}

Collins obtained the same formula in another way [17]. He considered particle
of mass $ m _ 0 $, located indefinitely far from a point source of a
gravitational field of mass $ M $. Such particle has energy
$ E _ 0 = m _ 0c ^ 2 $.
At movement on a distance $ r $ from a source of a field, particles energy
will increase up to size $ E = m _ 0c ^ 2 + (\gamma m _ 0M) /r $.
Collins offers to interpret this change of energy as change of a rest mass in
a gravitational field.

\begin{equation}
m\;=\;m_0\left(1+\frac{\gamma M}{rc^2}\right).
\label{bb21a}
\end{equation}
Then he uses a conservation law of a momentum $ mv = m _ 0v _ 0 $ and
receives the law of change of speed in a gravitational field
\begin{equation}
V \; = \; v _ 0\left (1 + \frac {\gamma M} {rc ^ 2} \right) ^ {-1}.
\label{bb22a}
\end{equation}
Supposing, that this law is distributed also to photons, we receive the
formula for change of photons speed in a gravitational field

\begin{equation}
v\;=\;c\left(1+\frac{\gamma M}{rc^2}\right)^{-1}
\;\approx\;c\left(1-\frac{\gamma M}{rc^2}\right).
\label{bb23a}
\end{equation}

It is possible to interpret the formulas (20), (23) as hit of a photon in
medium with a refraction index

\begin{equation}
n_2(r)\;=\;1+\frac{\gamma M}{rc^2}.
\label{bb24a}
\end{equation}

In that case, when the speed of a particle $ v $ is comparable to the speed
of light $ c $, in the formula (21) it is necessary to take into account
relativistic correction to a rest-mass $ m $ and to record it as
\begin{equation}
M \; = \; m _ 0\left (1 + \frac {\gamma M} {rc ^ 2} + \frac12\frac {v ^ 2}
{c ^ 2} \right).
\label{bb25a}
\end{equation}
Appropriate refraction index will look like
\begin{equation}
n'_2(r)\;=\;1+\frac{\gamma M}{rc^2} + \frac12\frac{v^2}{c^2}.
\label{bb26a}
\end{equation}

Such difference in definition of refraction index of a gravitational field is
connected with that the speech in these cases goes about different objects,
which differently interact with a gravitational field.
In ESM to these situations there corresponds also different rotations in
extended space.

\section{Gravitational effects in ESM}

1) Speed of escape.

Speed of escape $ v _ 2 $ is that speed, which should be given to a body
located on a surface of the Earth, that it could be deleted from Earth on an
indefinitely large distance.
Let $ M $ - mass of the Earth, $ m $ - mass of a body located at the Earth
surface, and $ R $ - radius of this surface.
The expression for the speed of escape is [18]
\begin{equation}
v_2\;=\;\sqrt{2gR}\;=\;\sqrt{\frac {2\gamma M}R}.
\label{bb27a}
\end{equation}

We will receive now formula (27) using ESM methods.

Let's consider a massive particle at rest, which removed to infinite large
distance from the Earth. Within the framework of our model such particle is
described by isotropic 5-vector of energy-momentum-mass (5).
Space motion in gravitational field along an axis $ X $ can compare movement
in extended space $ G (1,4) $ in a plane $ XS $ from a point with refraction
index $ n = 1 $ to point with refraction index $ n (r) $. Such motion is
described by a rotation (13).

Here rotation angle $ \psi $ express through refraction index $ n $.
Thus the massive particle at rest acquires speed

\[v\;=\;c\frac{\sqrt{n^2-1}}n.\]
As to in this case we consider a motion of a massive body, we assume natural
to use refraction index $ n _ 2 $.
Assuming, that it is close to unit, i.e. that

\begin{equation}
1\;\gg\;\varepsilon\;=\;\frac{\gamma M}{rc^2},
\label{bb28a}
\end{equation}
we receive, that

\begin{equation}
v\;\approx\;c\sqrt{2\varepsilon}.
\label{bb29a}
\end{equation}
In case, when $ r = R $ - radius of the Earth, the formula (29) coincides
with the formula (27) and gives the speed of escape $ v = v _ 2 $.

2) Red shift.

Gravitational red shift usually considered as a change of frequency of a
photon in the case of changing of a gravitational field, in which photon is
merged. In particular, at decreasing of strength of a field the frequency of
a photon also decreases, that is it reddens [14]. However Okun offers to
recognize that not frequency varies but varies wavelength of a photon, and
just it to name as red displacement [9,10]. Under our judgment both cases
are possible, but they corresponds to different physical situations and are
described by different
rotation angle $ \psi $ express through refraction index $ n $.

In a general theory of relativity the formula that describes change of light
frequency is [14]
\begin{equation}
 \omega\;=\;\frac{\omega_0}{\sqrt{g_{00}}}\;\approx\;
\omega_0\left(1+\frac{\gamma M}{rc^2}\right).
\label{bb30a}
\end{equation}
Here $ \omega _ 0 $ - frequency of a photon measured in universal time, it
remains constant at propagation of a beam of light. And $ \omega $ -
frequency of the same photon which measured in its own time. This frequency
is various in various points of space. If the photon was emitted by a massive
star, near to a star at small $ r $ the frequency of a photon is more, than
far from it at large $ r $. On infinity in the flat space, where there is no
gravitational field, the universal time coincides with own and $ \omega _ 0 $
there is an observable frequency of a photon.

Let's consider now same problem from the point of view of ESM.

Within the framework of our model to a photon located in blank space, the
isotropic 5-vector (4) is compared.
Process of its movement to the point with refraction index $ n $, at which
the change it of frequency happens, so also of energy, is described by a
rotation
in $ (TS) $ - plane. At these rotations the photon vector will be transformed                                                                                                                                                                                                                                                                                                                                                                                                                                                                                                                                                                                                                                                                                                                                                                                                                                                                                                                                                                                                                se, when $ r = R $ - radius of the Earth, the formula (29) coincides
										                                                                                                                                                                                                                                                                                                                                                                                                                                                                                   the formula (27) and gives the speed of escape $ v = v _ 2 $.
From here it is visible, that $ \omega _ 0 $ - frequency of a photon in
vacuum and $ \omega $ - it frequency in a field are connected by a ratio                                                                                                                                                                                                                                                                                                                                                                                                                                                                                           d shift.

\begin{equation}                                                                                                                                                                                                                                                                                                                                                                                                                                                                                                                                                   tational red shift usually considered as a change of frequency of a photon in the case of changing of a gravitational field, in which photon is merged. In particular, at decreasing of strength of a field the frequency of a photon also decreases, that is it reddens [14]. However Okun offers to recognize that not frequency varies but varies wavelength of a photon, and just it to name as red displacement [9,10]. Under our judgment both cases are possible, but they corresponds to different physical situations and are described by different s
\omega\;=\;\omega_0\cosh\theta\;=\;\omega_0 n.
\label{bb31a}
\end{equation}

We assume, that at calculation of change of photon frequency it is necessary
to use refraction index $ n _ 2 $ , as to the index of refraction $ n _ 1 $
was found in the supposition, that this frequency does not vary.
Substituting (24) in (31), we receive the formula
\begin{equation}
 \omega \; = \;\omega _ 0 n _ 2 \; = \;\omega _ 0\left (1 + \frac {\gamma M} {rc ^ 2} \right).
\label{bb32a}
\end{equation}
which coincides the formula (30). Thus in extended space model for red shift
is received the same expression, as in general theory of relativity.

In papers [15,16] Okun has offered to consider red shift of a photon as
change it of speed, momentum and wavelength, but the frequency was assumed
constant. He proceeded from a dispersing ratio for a photon with zero mass
in space with the Schwarzchild metric
\[G ^ {00} p _ 0p _ 0 \;- \; g ^ {rr} p _ rp _ r \; = \; 0.\]
The Schwarzchild metric is
\begin{equation}
G ^ {00} \; = \; (1-\frac {r _ g} r) ^ {-1}, \; \; g ^ {rr}=
(1-\frac {r _ g} r), \; \;
R _ g \; = \;\frac {2\gamma M} {c ^ 2}.
\label{bb33a}
\end{equation}
Assuming, that $ áp _ 0 = \hbar\omega = const $, the Okun has received for
relation of a momentum $ p _ r $ from a radius $ r $ expression
\begin{equation}
p_r(r)\;=\;\frac{\hbar\omega}á(1-\frac{r_g}r)^{-1}\;=\;p_r(\infty)n_1,
\label{bb34a}
\end{equation}
Where $ p _ r (\infty) $ - is momentum of the photon at infinity, where the
influence of gravitational field is absent.

Using connection between a momentum of a photon and its wavelength
$ \lambda (r) $, we receive expression
\begin{equation}
\lambda (r) \; = \; \frac {2\pi} {\omega} v=
\frac {2\pi c} {\omega} (1-\frac {r _ g} r) \; =
\; \frac {2\pi c} {\omega {n _ 1}} \; = \; \frac {\lambda (\infty)} {n _ 1}.
\label{bb35a}
\end{equation}

For speed of a photon $ v (r) $ the Okun receives expression

\begin{equation}
v(r)\;=\;\frac{\lambda(r)\omega}{2\pi}\;=\;c(1-\frac{r_g}r)\;=\;\frac c{n_1}.
\label{bb36a}
\end{equation}

Let's look now at transformations (34) - (36) from the ESM point of view.

As the frequency of a photon remains constant, but vary its momentum and mass
the appropriate transformation must be described by a rotation in the plane
$ (XS) $ of the spaces $ G (1,4) $. As the frequency does not vary,
we take refraction index $ n _ 1 $. At such rotation the speed of a photon
varies according to the formula

\begin{equation}
v\;=\;c\cos\psi\;=\;\frac cn_1.
\label{bb37a}
\end{equation}
This formula coincides the formula (36) for transformation of speed.
Being repelled from it is possible to receive the formula (35),
assigning change of a wavelength of a photon, when photon hit in a
gravitational field. As if to the formula (34) naturally we can not to
receive it, as to we should use at calculations not a dispersing ratio (32),
but curved in (1 + 4)-dimensional Schwarzchild metric analog of a full
dispersing ratio of a photon in space $ G (1,4) $

\begin{equation}
E^2\;-\;c^2p^2_x\;-c^2p^2_y\;-\;c^2p^2_z\;-\;m^2c^4\;=\;0,
\label{bb38a}
\end{equation}

From a point of view of our model it is necessary to consider the formula (31) only as first approximation to an exact result. Let's estimate correction appropriate to that in this model the photon, hitting in area with $ n> 1 $ gains a nonzero mass. For this reason the part of a photon energy can be connected not to frequency, but with the mass.
Let's estimate magnitude of this energy for case, when photon frequency change, in case of incident from a height $ H $ in a homogeneous gravitational field, with acceleration of gravity $ g $ is measured. Such situation was realized in well known Pound and Rebka experiments [19]. The energy change which appropriated to such frequency shift, is equal
\begin{equation}
\Delta E \; = \;\left (\frac {\hbar\omega} {c ^ 2} \right) gH.
\label{bb39a}
\end{equation}
According to the formula (6) in the case of rotation in plane $ (TS) $ the
photon gains a mass

\begin{equation}
m\;=\;\frac{\hbar\omega}{c^2}\sqrt{n^2-1}.
\label{bb40a}
\end{equation}
The difference of potential energies in the point of emission and point of absorption of a photon, which differ by height $ H $, is equal

\begin{equation}
\delta E\;=\;mgH\;=\;\left(\frac{\hbar\omega}{c^2}\right)gH\sqrt{n^2-1}.
\label{bb41a}
\end{equation}
Near to a surface of the Earth refraction index of gravitational field is define by the formula (16). Taking into consideration an inequality (28), we shall receive an evaluation

\begin{equation}
\delta E\;=\;mgH\;=\;\left(\frac{\hbar\omega}{c^2}\right)gH\sqrt{n^2-1}
\;\approx\;
\label{bb42a}
\end{equation}
\[\left(\frac{\hbar\omega}{c^2}\right)gH\sqrt{\frac{2\gamma M}{Rc^2}}
\;=\;\left(\frac{\hbar\omega}{c^2}\right)gH\sqrt{\frac{2gR}{c^2}}\;\approx\;
\left(\frac{\hbar\omega}{c^2}\right)gH(2.5\cdot 10^{-5}).\]
We see, that correction to effect connected to emerging of the photons
nonzero mass, near to the Earth surface is only $ 10 ^ {-5} $ from magnitude
of the total effect.

3) Delay of radar echo.

The appearance of radar echo delay is, that the time of light distribution up to some object, and back, can differ in dependence from that, does this light spread in a hollow, or in a gravitational field.
Such delay was measured in experiments on location of Mercury and Venus [20]. Such experiments give satisfactory agreement with GR predictions. These experiments also were analyzed in [21]. Here we do not interesting to analysis of these work. We want only to indicate that the analytical expression for magnitude of delay of a radar echo in ESM coincides what is received in GR.

This result can be obtained from the fact that the photon time delay $ \Delta t $ is calculated from only from the photon velocity $ v (t) $ [15,16]. Let's imagine that we locate the Sun. In this case we have
\begin{equation}
\Delta t \; = \; 2\left (\int\limits _ {R _ s} ^ {r _ e} \frac {dr} {v (r)} -
\int\limits _ {R _ s} ^ {r _ e} \frac {dr} c\right).
\label{bb43a}
\end{equation}
Here $ R _ s $ - radius of the Sun, $ r _ g $ - gravitational radius of the
Sun, and $ r _ e $ - distance from the Earth up to the Sun.

Speed of light in a gravitational field is $ v = \frac cn $. As here we deal
with photons, as a refraction index it is necessary to select $ n = n _ 1 $.
Substituting it in (43), we obtain

\begin{equation}
\Delta t\;=\;\frac{4\gamma M}{rc^2}\ln\frac{r_e}{R_s}.
\label{bb44a}
\end{equation}
The formula (44) coincides with expression for magnitude of radar echo delay
obtained in work [16,21].

4) In general theory of relativity magnitude of deviation angle
$ \delta\psi $ of a light beam from a rectilinear trajectory in case of
photon motion near to a massive body determine, deciding the eyconal equation
which defining trajectory of this beam in a central-symmetrical gravitational
field [14]. In this case we receive the answer

\begin{equation}
\psi\;=\;4\frac{\gamma M}{Rc^2}.
\label{bb45a}
\end{equation}
Here $ M $ - mass of a body, and $ R $ - distance at which the light beam
passes from a field center.

As in this case speech goes about photons motion it is necessary to select
$ n = n _ 1 $. Let's consider two beams - one passes precisely through an
edge of the Sun, and other at a distance $ h $ from it. It is supposed, that
$ h\ll R _ s < r $. In case of passing by these rays of a linear segment of
length $ dx $ the residual of optical paths will be

\begin{equation}
\delta x\;=\;dxn_1(r)-dxn_1(r+h\cos\varphi)\;=\;
\label{bb46a}
\end{equation}
\[dx\left(1-\frac{r_g}r\right)\;-\;dx\left(1-\frac{r_g}{r+h\cos\varphi}\right)
\;\approx\;\frac{r_gh\cos\varphi}{r^2}dx.\]
To such difference of optical paths there corresponds an angle of a wave
front deviation
\begin{equation}
\delta\varphi \;\approx \;\frac {\delta x}h=\frac{r_g\cos\varphi}{r^2} dx
\; = \; \frac {r _ gR _ s} {r ^ 3} dx \; = \;\frac {r _ gR _ sdx}
{(x ^ 2 + R _ s ^ 2) ^ {3/2}}.
\label{bb47a}
\end{equation}
Integrating this expression on $ x $ from $ -\infty $ up to $ + \infty $ we
shall receive deviation angle
\begin{equation}
\varphi \; = \; r _ g R _ s\int\limits _ {-\infty} ^ {\infty} \frac {dx}
{(x ^ 2 + R _ s ^ 2) ^ {3/2}}
\; = \; 2\frac {r _ g} {R _ s} \; = \; 4\frac {\gamma M} {R _ sc ^ 2},
\label{bb48a}
\end{equation}
Which coincides with an angle (45).

5) Perihelion precession of Mercury

 One more classical GR effect is the perihelion precession of Mercury.
It arises due to a space curvature the Newton's law of an attraction is
deformed. It reduces that the trajectory of a partiále becomes nonclosed, and
after of each rotation it the perihelion  precessed at some angle.
The magnitude of this rotation is determined by the law of interaction of a
central mass $ M $ and mass $ m $ particles rotated around it. In case of a
Schwarzchild potential the force of interaction of these masses is [22]

\begin{equation}
F(r)\;=\;-\frac{\gamma Mm}{r^2\sqrt{1-r_g/r-v^2/c^2}}.
\label{bb49a}
\end{equation}
Here $ r _ g $ - gravitational radius of mass $ M $, and $ v $ - velocity of
moving at the orbit partiále with mass $ m $.
The Mercury velocity of motion at the orbit around the Sun is equal
approximately 48 km/s, that gives magnitude of the relativistic correction
$ v ^ 2/c ^ 2\approx5\times 10 ^ {-8} $.

The gravitational correction has magnitude $ r_g/r\approx5\times 10^{-8} $
and it is comparable with relativistic correction. It is possible to consider,
that in the formula (49) masses of a partiále $ m $ depend on a distance and
velocity. But as both these corrections are small, the total transformation
of mass $ m $ can be write as

\begin{equation}
m\;\;\to\;\;m\left(1+\frac{\gamma M}{rc^2} +
\frac12\frac{v^2}{c^2}\right).
\label{bb50a}
\end{equation}
The calculation with use of force (49) and including into account an
approximation (50) gives an value of perihelium precession of Mercury close
to observed. Let's consider now this case in ESM framework.

We have a partiále with a nonzero mass were in a gravitational field.
As the partiále has a nonzero mass, it is necessary to use the refraction
index $ n _ 2 $. This is similar to our way of estimation of the escape
velocity. As the relativistic correction in this case cannot be neglected, we
shall use an index of refraction $ n'_2 $, which is determined by the formula
(26). However, now "motion" of a partiále from area with refraction index
$n = 1 $ to the area with refraction index $ n'_2 $ does not defined by real
motion particle in space. Such "motion" determined by changing of the force
magnitude, operating on a partiále, i.e. this is the case of particle energy
modification. Therefore in this case it is necessary to use a (TS) rotation in
extended space
$G(1,4)$.In the case of such rotation the massive vector (5) will be
transformed in accordance with (8).
Thus in case of such rotation the massive partiále changes the mass
\begin{equation}
m\;\;\to\;\;m   e^{\theta}\;,\;\;\;0\;\le\;\theta\;<\;\infty\;
\label{bb51a}
\end{equation}
That fact, that in the formula (8) both signs have a direct physical sense,
means, that in case of such transformation from one partiále with a mass $m$
there can origin two partiáles with different masses.

\begin{equation}
m\;\;\to\;\;m_+\;=\;me^{\theta_{+}}\;=\;m(n + \sqrt{n^2-1}).
\label{bb52a}
\end{equation}
\begin{equation}
m\;\;\to\;\;m_-\;=\;me^{\theta_{-}}\;=\;m(n - \sqrt{n^2-1}).
\label{bb53a}
\end{equation}

We shall assume, that in a macroscopic massive body there is an equal number
of partiáles which converts under the laws (52), (53), and consequently we
shall use the average law of transformation
\begin{equation}
M\;\;\to\;\;mn_2\;=\;m\left(1+\frac{\gamma M}{rc^2}\right).
\label{bb54a}
\end{equation}

We see, that the formula (54) coincides the formula (50).

\section{Discussion}

We have shown, that the GR predictions can be received, based on analogies
between optical and mechanical phenomena. For this purpose it is enough to
use only technique of rotations in extended space $ G (1,4) $ and formula for the
refraction index in the Schwarzchild metric. Any additional ideas and
suppositions were not attracted. Actually obtained results are only first
approximation in an evaluation of magnitude of gravitational effects.

These results can be improved, taking into account physical processes, which
happen in extended space. So for example, calculation of magnitude of light
deviation it was not taken into account, that the photon in a gravitational
field gains a mass, and on it affected additional force of an attraction.
However, it seems, that the magnitude of these additional effects is
insignificant. This is showing an example of an evaluation of correction (32)
to the frequency of a photon in the gravitational field.

However, the purpose of this work was not an evaluation of such corrections
and arguing of possibilities of their real observation. Our purpose was to
demonstrate that the ESM methods in the standard GR tasks give right results.
In future we assume to use this results in the tasks of cosmology for
description of dark energy and dark substance, which explanations have not
received yet. Therefore that fact, that in well-known situations ESM works
well. This gives hope that, and for other tasks model too will give right
results.

\vspace {2ex}

 LITERATURE

1. D.V. Sivuhin,  Optics,
M.: NAUKA, (1985).(in Russian)

2. D.Yu.Tsipenyuk, V.A.Andreev ,  Structure of Extended Space,
Preprint IOFAN (General Physics Institute) , 5, Moscow, 25p., (1999).

3. D.Yu.Tsipenyuk, V.A.Andreev ,  Electeodynamics in Extended Space,
Preprint IOFAN (General Physics Institute) , 9, Moscow, 26p., (1999).

4. D.Yu.Tsipenyuk, V.A.Andreev, Interaction in Extended Space,
Preprint IOFAN (General Physics Institute) , 2, Moscow, 25p., (2000).

5. D.Yu.Tsipenyuk, V.A.Andreev, Lienar-Vihert Potencials and Lorentz
Force in Extended Space,  Preprint IOFAN (General Physics Institute) ,
1, Moscow, 20p., (2001).

6. D.Yu.Tsipenyuk, V.A.Andreev, "Extended Space and the Model of Unified
Interacrtion",  Kratkie soobstcheniya po fizike (in Russian), N6, 23
(2000); (Bulletin of the Lebedev Physics Institute (Russian Academy of
Sciences), Alerton Press, Inc., N.Y.2000); arXiv:gr-qc/0106093, (2001).

7. D.Yu.Tsipenyuk, V.A Andreev, "Lienar-Vihert Potentials in Extended Space",
Kratkie soobstcheniya po fizike (in Russian), N6,p.3-15, (2002)
( Bulletin of the Lebedev Physics Institute (Russian Academy of Sciences),
Alerton Press, Inc., N.Y.2002);
arXiv: physics/0302006, (2003).

8. L.A. Rivlin,  Quantum Electronics, v.33, 777 (2003).

9. Yu.V. Baryshev, A.G. Gubanov, A.A. Raikov,  Gravitation, v.2, 72 (1996)
(in Russian).

10. V.V. Okorokov, Preprint ITEP, 27-98, Moscow, (1998).

11. V.N. Strel'tsov,  What Testifies the Experiments on the
Investigation of Gravitational Time Dilatation, Communication of the JINR,
 2-99-133, (1999) (in Russian).

12.  Kh.M. Beshtoev,   Defect Mass in Gravitational Field and
Red Shift of Atoms and Nuclear Spectra,  arXiv:quant-ph/0004074 v1,
(2000).

13. V.A. Dubrovskiy,  Measurements of the Gravity Wawes Velocity,
 arXiv:astro-ph/0106350, (2001).

14. L.D. Landau and E.M. Lifshits,  Classical Field Theory,
Pergamon Press (1961).

15. L.B. Okun, K.G. Selivanov, V.L. Telegdi,   Uspekhi Fiz. Nauk, 169,
1141 (1999)(in Russian).

16. L.B. Okun,  arXiv:hep-ph/0010120 v2, (2000).

17. R.L. Collins,  Gravity slows the speed of light,
 APS eprint server, (8/9/97).

18. I.I. Ol'hovskii,  Theoretical Mechanics for Physisists,
Moscow, NAUKA, (1970) (in Russian).

19. R.V. Pound, G. A. Rebka,  Phys. Rev. Lett., v.4, 337
(1960).

20. I.I. Shapiro, Phys. Rev. Lett., v.13, 789 (1964).

21. S. Weinberg, Gravitation and Cosmology, Wiley, (1972).

22. C. Moller, The Theory of Relativity, CLARENDON PRESS
OXFORD, (1972).

  \end{document}